\documentclass[aps,twocolumn,amsfonts]{revtex4}
\begin{document}
\newcommand{\bb}{\begin{equation}}
\newcommand{\ee}{\end{equation}}
\newcommand{\eqb}{\begin{eqnarray}}
\newcommand{\eqf}{\end{eqnarray}}

\preprint{}
\title{Vortices, Infrared effects and  Lorentz Invariance Violation }
\author{H. Falomir}
\email{falomir@fisica.unlp.edu.ar} \affiliation{IFLP -
Departamento de F{\'\i}sica, Facultad de Ciencias Exactas, Universidad
Nacional de la Plata, C.C. 67, (1900) La Plata, Argentina}
\author{J. Gamboa}
\email{jgamboa@lauca.usach.cl} \affiliation{Departamento de
F\'{\i}sica, Universidad de Santiago de Chile, Casilla 307,
Santiago 2, Chile}
\author{J.  L\'opez-Sarri\'on}
\email{justo@dftuz.unizar.es} \affiliation{Departamento de
F\'{\i}sica, Universidad de Santiago de Chile, Casilla 307,
Santiago 2, Chile}
\author{F. M\'endez}
\email{fernando.mendez@lngs.infn.it} \affiliation{INFN,
Laboratorio Nazionali del Gran Sasso, SS, 17bis,
  67010 Asergi (L'Aquila),  Italy
  \\ and
  \\
 Departamento de F\'{\i}sica, Universidad de Santiago de Chile,
Casilla 307, Santiago 2, Chile}
\author{A. J. da Silva}
\email{ajsilva@fma.if.usp.br} \affiliation{Instituto de F{\'\i}sica,
Universidade de S\~ao Paulo,
 CP  66316, 05315-970, S\~ao Paulo, SP, Brazil}

\begin{abstract}
The Yang-Mills theory with noncommutative fields is constructed
following Hamiltonian and Lagrangian methods. This modification of
the standard Yang-Mills theory produces spatially localized
solutions very similar to those of the standard non Abelian gauge
theories.  This modification of the Yang-Mills theory contain in
addition to the standard contribution,  the term $\theta^\mu
\epsilon_{\mu \nu \rho \lambda} \left(A^\nu F^{\rho \lambda} +
\frac{2}{3} A_\nu A_\rho A_\lambda\right)$ where $\theta_\mu$ is a
given  space-like constant vector with canonical dimension of
energy.  The $A_\mu$ field rescaling and the choice
$\theta_\mu=(0,0,0,\theta)$, suggest  the equivalence between the
Yang-Mills-Chern-Simons theory in 2+1 dimensions and QCD in 3+1
dimensions in the heavy fermionic excitations limit. Thus, the
Yang-Mills-Chern-Simons theory in 2+1 dimensions could be a
codified way to $\mbox{QCD}$  with only heavy quarks. The
classical solutions of the modified Yang-Mills theory for the
$SU(2)$ gauge group  are explicitly studied.
\end{abstract}
\maketitle

\section{Introduction}

It has been shown that noncommutative geometry is an important
mathematical ingredient that could be a clue for several important
unsolved problems in theoretical physics \cite{szabo}.  One of the
consequences of noncommutative geometry is the Lorentz invariance
symmetry breaking that, as was pointed out by several authors,
could happen at very high or very low energies as a consequence of
the IR/UV property, implying that interesting new phenomenological
possibilities could appear \cite{carroll,gural} and new possible
extensions of the standard model \cite{stand}.

In Ref.\ \cite{cortes}, an approach to a Lorentz invariance
violating quantum field theory has been proposed, inspired in
noncommutative geometry, where the fields (instead of satisfying
the standard canonical commutators) obey \eqb \left[\phi_i ({\vec
x}), \phi_j({\vec y})\right] &=& i \theta_{ij} \delta ({\vec
x}-{\vec y}), \label{c1}
\\
\left[\pi_i ({\vec x}), \pi_j({\vec y})\right] &=& i { B}_{ij}
\delta ({\vec x}-{\vec y}), \label{c2}
\\
\left[\phi_i ({\vec x}), \pi_j({\vec y})\right] &=& i \delta_{ij}
\delta ({\vec x}-{\vec y}), \label{c3} \eqf where $i,j,...=1, 2,3,
...$ are internal indices and $\theta$ and ${ B}$ are scales with
dimensions of $(\mbox{energy})^{-1}$ and energy respectively.
These scales correspond to ultraviolet and infrared weak Lorentz
invariance violations respectively.

These small deviations of the Lorentz symmetry (ultraviolet and
infrared) imply modifications to the relativity principle.  In the
ultraviolet sector \cite{amelino}, for example, it     allows to
describe an interesting phenomenology for UHECR,  where possible
new effects could be studied \cite{torres}.

The present approach does not correspond to the noncommutative
geometry in the true sense, where one adopts the commutator
 $$
    \left[x,y\right] \sim \theta
    .$$

Rather, while the commutator (\ref{c1}) violates the
microcausality principle imposing an ultraviolet scale, (\ref{c2})
affects the physics in the infrared sector. This is relevant in
the infrared sector of quantum field theory \cite{kost}, where
other phenomena could be explained.

Indeed, in the infrared region new windows could emerge to largely
unsolved problems, such as the dark matter and energy puzzle
\cite{kolb}, matter-antimatter asymmetry \cite{dine}, primordial
magnetic fields \cite{rubi} and other interesting phenomena.
However, important open questions concerning the meaning of the
infrared scales are still unsolved  \cite{pepe}.

Although  there are no definitive general answers to these
problems, one can consider particular examples which  eventually
could be confronted with phenomenology or experimental results.

The purpose of the present research is to extend our previous work
for electrodynamics \cite{us2} to the non-abelian case. In
\cite{us2} we shown that, by deforming the canonical algebra in
the infrared sector, one rediscover the Carroll-Field-Jackiw
theory proposed fifteen years ago \cite{cfj}.

Although in the context of our approach we do not know how  to
treat fermionic degrees of freedom, we will consider that this
modified Yang-Mills theory hide the fermions in a Chern-Simons
term. From this point of view, it seems to be reasonable to think
that confinement could be a phenomenon mainly due to the classical
behavior of the gluonic  fields \footnote{In another context, this
problem was studied in \cite{kj}.}.

The aims of the present paper are the following:

1) To present a nonabelian gauge field theory that can be
understood as a Hamiltonian system where the commutators --or
Poisson brackets at the classical level-- are deformed in a
similar way to those of a noncommutative system (although we
strength that this noncommutativity is in the field space, not in
the spacetime as in noncommutative geometry),

2)  To show that our approach can also be understood  as a
standard Yang-Mills Lagrangian plus the term

\bb \theta^\mu \epsilon_{\mu \nu \rho \lambda} \left(A^\nu F^{\rho
\lambda} + \frac{2}{3} A^\nu A^\rho A^\lambda\right), \label{cs21}
 \ee
where vectorial fields $A_\nu$ satisfy canonical commutations
relations and $\theta_\mu$ is a given  space-like vector having
dimensions of $\mbox{energy}$. One should also emphasize that this
Chern-Simons term not only violate Lorentz invariance, but also C,
P and T symmetries \cite{kost2}.

3) To  discuss how the modified Yang-Mills equations in four
dimensions can be understood in three-dimensions as a
Yang-Mills-Chern-Simons system and how the behavior of the gauge
fields is very similar to the Nielsen-Olesen vortices.

4) To give arguments that suggest that this modified
Yang-Mills-Chern-Simons theory could be understood as a kind of
\lq \lq bosonized" $QCD$ theory at low energy.

\section{Yang-Mills theory with noncommutative fields}\label{int:2}

In this section we will construct the Yang-Mills theory with
noncommutative fields following similar arguments to those given
in ref. \cite{us2} (in other context see also ref.
\cite{andrianov}). In this paper, Latin indices denote spatial
components and the metric is taken as diag$(-1,1,1,1)$.

Essentially, one starts considering the following modified Poisson
brackets
\begin{eqnarray}
\{A^a_i(x),A_j^b(x') \}_{P.B} &=& 0, \label{1}\\
\{A^a_i(x),\Pi_j^b(x') \}_{P.B} &=&\delta^{ab}
\delta_{ij}\delta^3(x-x'),  \label{2}\\
\{\Pi^a_i(x),\Pi_j^b(x') \}_{P.B} &=&
\epsilon_{ijk}\theta^k\delta^{ab} \delta^3(x-x'),   \label{3}
\end{eqnarray}
where the parameter $\theta_k$ ($k= 1,2,3$) is a vector in the
space that is responsible for the Lorentz invariance violation.
Also, the indices $a,b, c, ...$ represent internal indices,
corresponding to the gauge group. In our analysis we will take the
$SU(2)$ group as an example whose structure constants are just
$i\epsilon^{abc}$, {\it i.e.} the total antisymmetric tensor. One
should note that a term like $\epsilon^{abc}\gamma_c
\delta_{ij}\delta^3 (x-x')$  could also be added to the right hand
side of (\ref{3}), but  since we are looking for terms which
violate the Lorentz symmetry, this contribution would be
irrelevant in the following analysis.

Following  \cite{us2}, we will  keep the gauge symmetry exact
\footnote{This condition allows us to discard in (\ref{3}) the
term previously  mentioned, otherwise we would be forced to
include it.}  while breaking the Lorentz invariance. Therefore,
the first step is to find in this context the correct generators
for the gauge transformations in terms of $A$ and $\Pi$ fields.

We then modify the standard Gauss law,
\begin{equation}
\Psi_a(x) \equiv \vec\nabla\cdot\vec\Pi_a +
g\epsilon^{abc}\vec\Pi_b \cdot \vec A_c = 0, \label{oldgauss}
\end{equation}
by adding a new term, $\Lambda(x)$, to get

\bb \Psi'(x)=\Psi(x)+\Lambda(x). \label{new} \ee

This new term should depend only on the gauge potentials in order
to reproduce  the usual gauge transformation on $A$.

Taking into account the commutation relations

\eqb \{\Pi^a_i(x)&,&B^b_k(x')\} = \nonumber
\\
&-&\epsilon_{ijk}\biggl( \delta^{ab}\partial^j \delta^3(x- x') +
g\, \epsilon^{abc}A_c^j(x)\delta^3(x-x')\biggr), \nonumber \eqf

and

\eqb \{\Pi^a_i(x)&&,\Psi^b(x)\}=
-g\,\epsilon^{abc}\Pi_{i\,c}\delta^3(x-x') \nonumber
\\
-&&
\epsilon_{ijk}\theta^k\left(\partial^j\delta^3(x-x')\delta^{ab} +
g\,\epsilon^{abc} A_c^j \delta^3(x-x') \right), \nonumber
\\
\label{ww} \eqf

where $B^b_k = \frac {1}{2} \epsilon_{i j k} F^{b \,i j}$, one
finds that the commutator between $\Pi_i^a$ and $\Psi^a(x)$  is
given by
\[
\{\Pi_i^a(x),\Psi^b(x') -\vec\theta\cdot\vec B^b(x')\} =
-g\,\epsilon^{abc} \Pi_{c i} (x)\delta^3(x-x').
\]

Therefore,  the correct generator of the gauge transformations in
this theory becomes

\begin{equation}
\Xi^\prime_\Omega = -\int
d^3x\,\Omega^a(x)\left(\vec\nabla\cdot\vec\Pi^a +
g\,\epsilon^{abc}\vec\Pi_b\cdot\vec A_c - \vec\theta\cdot\vec
B^a\right),
\end{equation}
or equivalently

\begin{equation}
\Xi^\prime_\Omega= -\int d^3x\,\Omega^a(x)\left((\vec
D\cdot\vec\Pi)^a -\vec\theta\cdot B^a\right),
\end{equation}
where $\vec D$ is the covariant gradient.

Similarly to the commutative case, one can propose the Hamiltonian

\eqb H &=&\int d^3x\,\frac{1}{2}\left(\vec\Pi^a\cdot\vec\Pi^a +
\vec B^a\cdot \vec B^a\right) \nonumber
\\
&+& \int d^3x\,A^{a\,0}\left(\vec\nabla\cdot\vec\Pi^a +
g\,\epsilon^{abc}\vec\Pi^b\cdot\vec A^c-\vec\theta\cdot\vec B^a
\right).  \nonumber
\\
\label{hamilton1} \eqf

Therefore, the equations of motion becomes \eqb \vec\Pi^a &=&
\dot{\vec A^a} + \vec\nabla A^{a\,0} + g\,\epsilon^{abc}
A_{b\,0}\vec A_c\, , \nonumber \label{1ee}
\\
\{\Pi_i^a(x),H\} &=& \epsilon_{ijk}(D^j B^k)^a - g\,\epsilon^{abc}
A^{b\,0} \Pi_c^i - \vec\theta\times\vec\Pi^a\, . \nonumber
\\
\label{2ee} \eqf The first equation establishes that $\Pi^a_i=
-F^{a\,0}_i$. The second equation can be written as
$$
(D_0\Pi_i)^a = \epsilon_{ijk}(D^j B^k)^a
-\vec\theta\times\vec\Pi^a,
$$
where the last term introduces a modification with respect to the
commutative case.

\smallskip

Finally, one can  find an equivalent Lagrangian which reproduces
the same equations of motion by identifying  a set of commuting
fields $\mathcal{P}$ \cite{mar} such that the $A$'s and
$\mathcal{P}$'s be canonically conjugate variables. These linear
combination of $\Pi$'s and $A$'s can be identified as
\[
{ P}^a_i \equiv \Pi^a_i -\frac{1}{2}\epsilon_{ijk}A^{j a}\theta^k,
\]
which satisfy the canonical algebra
\begin{eqnarray}
\{ A_i^a(x),{P}_j^b(x')\}&=& \delta_{ij}\delta^{ab}\delta^3(x-x')
\\
\{{P}_i^a(x),{P}_j^b(x')\}&=&0.
\end{eqnarray}

In terms of these new variables the Lagrangian becomes
\[
L=\int d^3x\,{ P}^a_i\dot A^a_i - H,
\]
where the only  modifications  comes from the terms proportional
to $\vec\theta$.

Thus, the equivalent Lagrangian density becomes
\begin{equation}
{ L}= { L}_0 + \frac{1}{2}\, \theta_k C_k, \label{lag}
\end{equation}
where ${ L}_0$ is the Yang-Mills Lagrangian density,
\[
\mathcal{L}_0 = -\frac{1}{4} \, F^a_{\mu\nu} F^{a\,\mu\nu},
\]
with
\[
F_{\mu\nu} = \partial_\mu A_\nu - \partial_\nu A_\mu + g \left[
A_\mu,A_\nu \right],
\]
and $C_k$ given by,
\begin{equation}
C_k= -\epsilon_{ijk}\left(- \dot A^a_i\,A^a_j +
A^{a\,0}F^a_{ij}\right).
\end{equation}

Using  $\dot A_i = \partial_0 A_i$ and the definition of
$F_{ij}^a$, one finds,

\eqb \int d^3x\, C_k &=&  \int d^3x\,\epsilon_{ijk}\biggl(-
A_i^a\partial_0 A^a_j - A_0^a\partial_i A^a_j +  A^a_0\partial_j
A^a_i \nonumber
\\
&-& g\,\epsilon^{abc} A^a_0 A^b_iA^c_j\biggr) \nonumber
\\
&=& \int d^3x\,2\,\epsilon_{k \nu  \rho \sigma} {\rm
tr}\left(A^\nu F^{\rho \sigma} + \frac{2}{3}g A^\nu A^\rho
A^\sigma \right). \nonumber
\\
\eqf

Collecting all the terms one finds that (\ref{lag}) can be written
as \bb { L} = - \frac{1}{2} {\rm tr} \left\{F_{\mu \nu}
F^{\mu\nu}\right\} + 2\,\theta^\mu \epsilon_{\mu \nu \rho \sigma}
{\rm tr}\left(A^\nu F^{\rho \sigma} + \frac{2}{3}g A^\nu A^\rho
A^\sigma \right), \label{lag2} \ee where the $A F$ and $A^3$
nonabelian contributions coincides with the Chern-Simons term and
$\theta^\mu$ is a space-like vector \cite{andrianov}.

Thus, (\ref{lag2}) shows the equivalence between a Hamiltonian
formulation with deformed commutators (or Poisson brackets) and a
standard Lagrangian formulation which  explicitly breaks down
Lorentz invariance.

Finally, we note that the Chern-Simons term should be treated as a
nonperturbative contribution. Indeed, if one rescales $A^a_\mu
\rightarrow g^{-1} A^a_\mu$ then the action becomes
\[
S=\int d^4 x\left[ \frac{1}{2 g^2} \mbox{tr} (F^2) +
\frac{\theta_\mu}{g^2}
  \epsilon^{\mu \nu \rho \sigma} \mbox{tr} \left(A_\nu F_{\rho \sigma}
  + \frac{2}{3} A_\nu A_\rho A_\sigma \right) \right].
\]
Therefore, the Chern-Simons term must be considered at the same
foot as the standard kinetic  $F^2$ part in the $g$ expansion.

As an example of this last fact, we think it is instructive to
check a simple case. Let us consider the plane wave solutions
given by Coleman \cite{coleman}.

 This solution of the Yang-Mills theory is given by the ansatz,

$$A_{+}^a = f^a(x^+)x^1 + g^a(x^+)x^2 + h^a(x^+).$$

Here we are using light-cone coordinates, $A_\pm^a=A_0^a\pm A_3^a$
and $x^\pm= x^0\pm x^3$. The functions $f^a$, $g^a$ and $h^a$ are
arbitrary but decreasing like $\vert x\vert^\alpha$, with $\alpha$
a negative constant, for large arguments $\vert x\vert$. So, the
strength tensor becomes $F_{+1}^a=f^a$ and $F_{+2}^a=g^a$.

Then, we can consider a correction of this solution depending on
$\theta$ perturbatively. This will give a perturbative ansatz for
$F$,
$$F^a_{\mu\nu} = F^{(0)a}_{\quad\mu\nu} + F^{(1)a}_{\quad\mu\nu} +
{ O}(\theta^2)$$ where $F^{(0)}$ is the Coleman solution and
$F^{(1)}$ is the first order correction in $\theta$. Then, the
equations of motion up to first order are, \eqb
\partial_\mu F^{(1)\,a\,\mu\nu} &+&
g\epsilon^{abc}(A^{(0)\,b}_{\quad\mu}F^{(1)\,c\,\mu\nu} \nonumber
\\
&+& A^{(1)\,b}_{\quad\mu} F^{(0)\,c\,\mu\nu}) -
\frac{1}{2}\theta_\mu\epsilon^{\mu\nu\sigma\rho} F^{(0)\,a}_{\quad
  \sigma\rho} = 0.\nonumber
  \\
\eqf

It is easy to see that, for large $\vert x^+\vert$, the
perturbation $F^{(1)}$ goes as $\vert x^+\vert^{\alpha+1}$ and
then, for large distances, it is bigger than the zero-th   order
contribution to the perturbative solution. Hence, it is not
justified to take the noncommutative contribution as a
perturbation  to the Yang-Mills equations.

In the next section, we will deal with an exact solution for the
complete (sourceless)  equations of motion on
$\mathbb{R}^2\backslash\{0\}$. We will find that it corresponds to
  nonperturbative vortex configurations.

\section{Vortex solutions of the Modified Yang-Mills Theory}\label{int:3}

Generally speaking one should note that the modification in
(\ref{lag2}) breaks rotational invariance, and the equations of
motion become equivalent to a coupled Yang-Mills-Chern-Simons
system if the space-like vector $\theta_\mu$ is chosen in a
particular spatial direction.

This last fact is quite interesting. Indeed, if one choose
$\theta_\mu = (0,0,0,\theta)$,  one finds an almost
Yang-Mills-Chern-Simons theory with noncommutative gauge fields in
$2+1$ dimensions after to use a suitable rescaling of fields. The
difference, however, is that the $A_\mu$ field depend on
$(x_0,x_1,x_2,x_3)$ instead of $(x_0,x_1,x_2)$ as usual. This
result seems to be completely general.

One should also note that, in analogy with the quantum Hall
effect, physical excitations like quarks in a Yang-Mills Lorentz
symmetry breaking theory necessarily must live in $2+1$-dimensions
(although the $A_\mu$ field is four-dimensional).

The above discussion can also be extended to any gauge group.

Now, we will discuss an exact solution for the noncommutative
$SU(2)$ Yang-Mills theory with a vortex behavior.

The modified Yang-Mills equations are, \bb \left( D_\nu F^{\mu
\nu}\right)^a -\frac{\theta_\nu}{2} \epsilon^{\mu \nu \rho \sigma}
F^a_{\rho \sigma} =0. \label{ym1} \ee

The solutions for these equations have been extensively discussed
in the literature (for a review see, ref. {\it e.g}  \cite{vinet})
and, in particular, the vortex-like solutions for
Yang-Mills-Chern-Simons are well known.

However, we emphasize here that, although  these vortex solutions
fit perfectly in our problem,  they are also mandatory if the
canonical commutators are modified as in (\ref{1})-(\ref{3}).
Indeed, the  $\theta$ parameter imply the choice of a  particular
plane and -- as we are interested in the infrared limit-- one
could neglect {\it mutatis mutandis} the short distances effects.

In order to solve (\ref{ym1}), let us consider a set of
coordinates in $\mathbb{R}^3$, and the unitary vectors in the
plane $x_3=$const. , ${\hat \phi}_i  = \epsilon_{ij}x^j/\rho,
\,\,\,\,{\hat \rho}_i = x_i/\rho$, where $\rho$ is the standard
radial coordinate in the plane. Let us consider the following
axially symmetric ansatz for the gauge fields

\bb A_0^a = {\hat \phi}^a \psi_2 (x_0,{\vec r}),
\,\,\,\,\,\,\,\,\,\, A^a_i = {\hat \phi}^a  {\hat \phi}_i \psi_1
(x_0,{\vec r}) + \delta_3^a {\hat \phi}_i \, \frac{1}{\rho},
\label{ans1} \ee and $A_3^0 = 0$.

Using  (\ref{ym1}), one finds that

\eqb \psi_1 &=&  c\, e^{- \alpha x_0} ~K_{1} (M \rho ),
\label{ppsi1}
\\
\psi^{'}_2 &=&  \theta ~\psi_1  \label{psi1}
 \eqf
where $\psi^{'}_2 = d\psi_2/d\rho$  and $K_{1}(x)$ is the Bessel
function of the second kind.

The coefficient $M$ is defined as
\[
M= \sqrt{ \theta^2 + \alpha^2},
\]
and $c$ and $\alpha$ are constant with dimensions of energy. The
gauge potential given in (\ref{ans1})  fall exponentially to zero
when $\rho \rightarrow \infty$.

For the configuration discussed above, the chromomagnetic energy
is finite \eqb { E}_m &=&\frac{1}{2} \int d^2x ~B_a \cdot B_a
\nonumber
\\
&=& \frac{\pi c^2}{2}. \label{electt} \eqf But the chromoelectric
energy is logarithmically divergent at the ultraviolet region:

\eqb { E}_e&=& \frac{1}{2}\int d^2x \vec E^a\cdot\vec E^a
\nonumber
\\
&=& \pi\int_\Lambda^\infty d\rho\,\rho
\left[(\Psi_2^\prime)^2+(\dot\Psi_1)^2\right], \label{magg}
\nonumber
\\
&=& \pi\,c^2  \int_\Lambda^\infty du\, u K_1^2(u), \eqf where
$\Lambda$ is a given cutoff. However, one should notice that it is
at the infrared region where $\theta$ is relevant.

Thus, we see that in four dimensions one finds that the energy of
these solutions increases linearly with $L$ for large distances
and, therefore, the gluonic fields would appear as confined along
the $z$ direction.

Finally, we would like to sketch  a possible origin of the \lq \lq
four-dimensional" Chern-Simons term.  In so doing, let us consider
massless QCD in four dimensions, described by the Lagrangian 
\bb 
{  L}= - \frac{1}{4}F^2 + {\bar \psi} \left( i D \hspace{-.6em}
\slash \hspace{.15em} \right) \psi, \label{palmita} \ee where a
sum over flavor indices is assumed.

Naively, one could expect that, by integrating the quark fields to
find an effective action at low energies, one could obtain
contributions  different from a Chern-Simons term. However, one
also can argue the following: at very  low energy, considering
only heavy quarks, let us suppose that the space-time is
compactified so that the quark field could be written as
\[
\psi (x_0,x_1,x_2,x_3) = e^{i\frac{x_3}{\ell}} \varphi
(x_0,x_1,x_2),
\]
where $\ell$ is the compactification radius \footnote{ This
compactification occur when we consider, for example, a membrane
vibrating in the space. If the transversal amplitude is small
enough, then the phonons propagate only on the surface of the
membrane, and the compactification is a good approximation.} ( for a more detailed discussion see \cite{cruz}).

Once this compactification is assumed,  the \lq \lq heavy quark"
acquires  an effective mass $m= 1/\ell$, and eq. (\ref{palmita})
in this effective description becomes 
\bb 
{ L}= - \frac{1}{4}F^2 +
{\bar \varphi} \left( i D \hspace{-.6em} \slash \hspace{.15em} - m
\right) \varphi, \label{palmita1} \ee where ${\bar \varphi} \left(
i D \hspace{-.6em} \slash \hspace{.15em} - m \right) \varphi$ is a
fermionic three-dimensional Lagrangian. Here, the fermionic
determinant can be calculated and the result at the lowest order
in $1/m=\ell$ is the Chern-Simons term \cite{redlich}.

It is worthwile to notice that this kind of topological terms
coming from integrated-out fermions appear also in different
contexts. For example, D'Hoker and Farhi consider fermions getting
large masses through Yukawa couplings to Higgs fields
\cite{chapata}, obtaining at low energies a Wess-Zumino-Witten
term as a relic of the quark degrees of freedom in the heavy mass
limit.

Another interesting point is how this 2+1-dimensional case is
related to the 3+1-dimensional one.  The answer to this question
is quite simple: the connection between the gauge field in three
and four dimensions is
\[
A^{(3)}_\mu \rightarrow \sqrt{\ell} A_\mu,
\]
then with this rescaling the four dimensional measure become
\[
\frac{d^4x}{\ell} \rightarrow  {\tilde d^3x}.
\]

We conclude this section emphasizing that the analysis here
presented could be a new route to understand some nonperturbative
aspects of QCD.

\section{ Conclusions }\label{int:5}

In this paper we have shown that  deforming the Poisson brackets
for the canonical momenta in a Yang-Mills theory, the resulting
theory is equivalent to a Yang-Mills-Chern-Simons system. The
classical theory --taking the SU(2) group--  has vortex like
solutions similar to the Nielsen-Olesen ones. The difference,
however, is that in our case they appear as a consequence of
Lorentz invariance violation.

However, we emphasize that our result does not imply that quarks
fields are absent in our approach. Rather, the Chern-Simons term
contains the information about the fermionic degrees of freedom
and, in this sense, our procedure could provide an alternative
route to study nonperturbative effects.

\bigskip

\vskip 0.5cm

We would like to thank C. Wotzasek for discussions.

This work was supported by grants from UNLP (grant 11/X381) and
CONICET (PIP 6160) (H.F.),  FONDECYT 1050114, DICYT-USACH  and
MECESUP 0108 (J.G and J. L.-S.) and by  Funda\c c\~ao de Amparo a
Pesquisa do Estado de S\~ao Paulo (FAPESP) and Conselho Nacional
de Desenvolvimento Cient\'{\i}fico e Tecnol\'ogico  (CNPq)
(A.J.S.).

\end{document}